# Effect of geometry on the frequency limit of GaAs/AlGaAs 2-Dimensional Electron Gas (2DEG) Hall effect sensors

Anand Lalwani[1], Miriam Giparakis[2], Kanika Arora[4], Avidesh Maharaj[1], Akash Levy[1], Gottfried Strasser[2], Aaron Maxwell Andrews[2], Helmut Köck[3], and Debbie G. Senesky[1,4]*

[1] Department of Electrical Engineering, Stanford University, Stanford, CA 94305, USA
[2] Institute for Solid State Electronics, TU Wien, Gußhausstraße 25,1040 Vienna, Austria
[3] Infineon Technologies Austria AG, Siemensstrasse 2, 9500 Villach, Austria
[4] Department of Aeronautics and Aerospace Engineering, Stanford University, Stanford, CA 94305, USA

* Member, IEEE

**Abstract**—In this work, we experimentally investigate the frequency limit of Hall effect sensor designs based on a 2-dimensional electron gas (2DEG) gallium arsenide/aluminum gallium arsenide (GaAs/AlGaAs) heterostructure. The frequency limit is measured and compared for four GaAs/AlGaAs Hall effect sensor designs where the Ohmic contact length (contact geometry) is varied across the four devices. By varying the geometry, the trade-off in sensitivity and frequency limit is explored and the underlying causes of the frequency limit from the resistance and capacitance perspective is investigated. Current spinning, the traditional method to remove offset noise, imposes a practical frequency limit on Hall effect sensors. The frequency limit of the Hall effect sensor, without current spinning, is significantly higher. Wide-frequency Hall effect sensors can measure currents in power electronics that operate at higher frequencies is one such application.

## I. INTRODUCTION

Hall effect sensors have been used extensively in industrial and research settings to quantify the magnetic field strength of a system in application-fields such as automotive [1], current sensing [2], and power electronics [3]. These sensors rely on Lorentz's force and the Hall effect. The voltage produced is directly proportional to the bias current input and the strength of the magnetic field. Due to non-idealities in fabrication, thermal gradients, and piezoelectric effects, all Hall effect sensors suffer from a non-zero offset voltage, which is a persistent noise in every Hall effect sensor and is typically removed using current-spinning techniques, which biases the Hall effect sensor from one pair of contacts to another, in effect changing the direction of the flow of the current while measuring the voltage on the remaining two contacts [4]. The average of the measured voltage is taken, thus canceling the offset. However, a small offset on the order of microvolts (μV) is still present.

Current spinning poses significant limitations on the speed at which a Hall effect sensor can be used. In steady-state or slowly oscillating magnetic fields, the current spinning switching frequency is significantly faster than the external magnetic field frequency to produce reliable results. However, when the magnetic field frequency exceeds the current spinning frequency limit, the current spinning method can no longer be used [5]. Most Hall effect sensors on the market cannot exceed 200 kHz [6], whereas power electronics such as DC-DC converters and inverters operate in the MHz region.

Recent work has shown that novel methods, such as the X-Hall architecture [5] or the 2-Omega [7] method, where the Hall effect sensor is biased using an AC current, remove the need for current spinning to reduce the offset noise and extend the operating frequency. There is limited literature on the practical frequency limit of Hall effect sensors, specifically for 2-dimensional electron gas (2DEG) heterostructures.

Popović [8] explains that the fundamental frequency limit of a Hall effect sensor is determined by the carrier relaxation time. However, a more practical limit is set by the resistance and capacitance (the RC constant) of the sensor. Crescentini et al. [9], have shown that the practical limit for Hall effect sensors can be determined by sending a pulse current across two nodes of the Hall effect sensor and measuring the transient voltage produced across the other two terminals. This measures the time taken for the offset voltage to rise in response to a pulsed bias current. They showed the equivalence between the simulations and experiments in order to measure the practical limit of silicon-based Hall effect sensors [9].

High mobility 2DEG heterostructures, like GaAs/AlGaAs, allow Hall effect sensors to operate at higher frequencies (up to MHz). This work expands on prior literature [10] to first measure the practical frequency limit of high-mobility GaAs/AlGaAs Hall effect sensors and second to study the role of the geometry of the Ohmic contacts and the trade-off between the frequency limit and sensitivity.

## II. FABRICATION AND EXPERIMENTAL SETUP

To conduct this experiment, device heterostructures were grown by molecular beam epitaxy (MBE) in a Riber C21 chamber on an undoped GaAs substrate. First, a 50 nm GaAs buffer layer and a 120 nm GaAs/AlAs short period smoothing superlattice (SPSL) were grown. To produce a 2DEG at the marked location in Fig 1(a), Si delta doping was introduced between the 20 nm thick $Al_{0.33}Ga_{0.67}As$ layer and the 40 nm thick AlAs/AlGaAs SL with an average Al fraction of 45%. This 2DEG design is optimized for stability in the entire temperature range rather than peak mobility.

The fabrication of the four Hall effect sensor designs, shown in Fig 1(b), was as follows: The active region mesa was defined by a Cl/Ar dry etching process that yielded smooth, slightly positively sloped sidewalls. For Ohmic contacts, Ge/Au/Ni/Au layers of 15/30/14/165 nm were sputtered and subsequently rapidly annealed at 450°C for 3 min. The device was passivated using a 200-nm-thick $Si_3N_4$ layer. The passivation was opened again on areas over the Ohmic contacts and the substrate and then Ti/Au contacts with a thickness of 10/220 nm were sputtered.

The four geometries of the Hall effect sensors were chosen for to study effect of geometry on sensitivities and signal to noise ratio [10, 11], as seen in Fig 1(c). The types 'long,' 'equal,' 'point,' and 'short' were co-fabricated on the same die, wire-bonded, and subsequently coated with epoxy. The characterization setup, shown in Fig 2, follows closely to reference [9]. The current is pulsed using a waveform generator (Agilent 33120A) with a square wave of 1 V, 5 ns edge rise time, and 50 kHz frequency. The pulse is applied to two opposite contacts and the voltage is measured using a differential amplifier (200 MHz Probe master) on the other two contacts and recorded at 5 giga samples per second with an oscilloscope (Keysight InfiniVision DSOX4022A). The Hall effect sensor is placed under a magnetic field of 63 mT to produce the Hall-voltage response. An average of 5 pulses is taken for each sensor and the following formula used to measure the rise time of the sensors.

$$V(t) = V_o e^{-t/\tau} \tag{1}$$

V(t) is the time-dependent voltage, $\tau$ is the rise-time, and $V_o$ is the base-line voltage. The time taken for the rise time $\tau$ is thus from 0 to 63.2% ($e^{-1}$) of the original value and is calculated. To convert rise time to bandwidth frequency, is multiplied to the inverse of the rise time [12]. The sensitivity of the sensors was measured using a Hall effect measurement system (LakeShore 8404) where the biased current was varied from 100 µA to 500 µA in increments of 100 µA and the magnetic field was a factor 0.35 varied from -0.9 T to 0.9 T in increments of 0.1 T. The voltage was measured in 6 current biased spin phases and the average value was taken to remove the offset. All measurements were performed at room temperature.

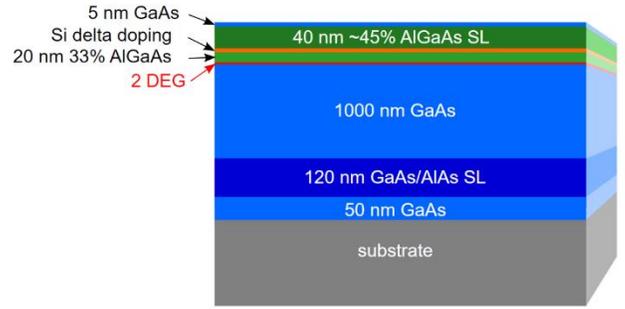

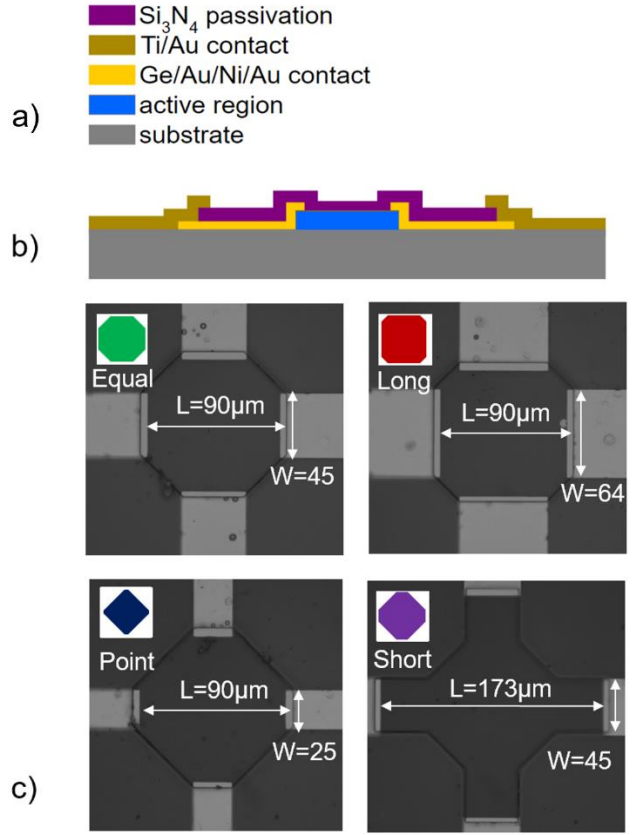

Fig. 1. The fabrication and geometries of the GaAs/AlGaAs Hall effect sensor. (a) the layer sequence as grown by MBE (not to scale). (b) a side-cut view of the fabricated Hall effect sensor (not to scale). (c) the four Hall effect sensor designs on the same die with the various geometries, names, and symbols used to denote them throughout this work.

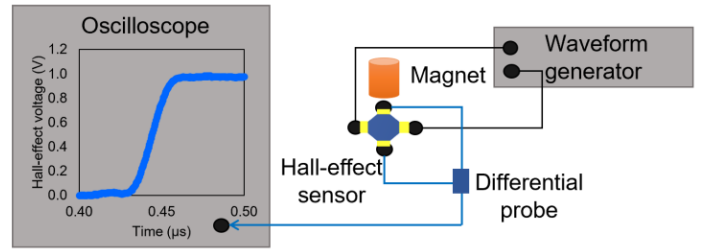

Fig. 2. Experimental setup for measuring the frequency limit of Hall effect sensors.

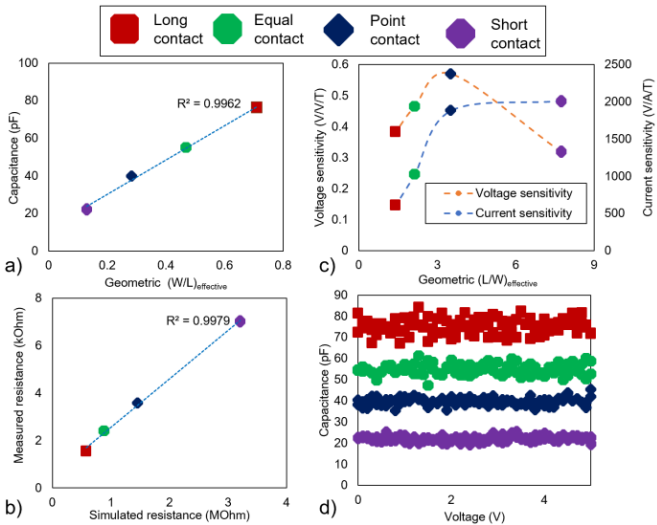

Fig. 3. The experimental investigation of the capacitance and sensitivity of the four Hall effect sensors. (a) The relationship between capacitance and the effective geometry (W/L). (b) The experimental and simulated resistance with a linear fit. (c) The voltage and current scaled sensitivity as a function of the geometric factor (L/W). (d) The capacitance-voltage plot for all 4 geometries.

Further investigation of the resistance and capacitance was conducted using a source meter (Keithley 2400) and a semiconductor device analyzer (Agilent B1500A). To find the effective geometric factor, since the sensor shapes are non-standard, finite element simulations were performed with COMSOL.

The resistance measurements for all four device geometries are a linear match with the resistance simulations from COMSOL as seen in Fig 3(b). Thus, we can calculate the effective geometric factor. The capacitance measurement at 1 kHz for all four geometries from 0 to 5 V is shown in Fig 3(d). The average of the capacitance values measured is in linear alignment with the effective width-to-length ratio, (W/L)$_{effective}$, and shown in Fig 3(a). As expected, the largest contact geometry device ('long') has the highest capacitance and similarly the device with the smallest W/L$_{effective}$ ratio has the lowest capacitance. The voltage-scaled sensitivity and current-scaled sensitivity for all four geometries is shown in Fig 3(c) and follow previously established trends [10].

## III. RESULTS AND DISCUSSION

The two primary factors that determine the frequency limit of a Hall effect sensor are the resistance and capacitance of the sensor. To vary these, the width of the Ohmic contacts of the Hall effect sensor is varied and the length between two Ohmic contacts in the active region is varied for all four sensors. The four sensors are on the same die, and thus have similar sheet resistance and material properties (contact resistance, electron mobility, and sheet carrier concentration). Further study is done to understand the trade-off between the frequency limit and the sensitivity of the devices.

Fig 4 shows the linear relationship between the geometric factor (L/W)$_{effective}$ and the resistance measured. As the length of the active region is increased from the 'equal' to 'short', we see an increase in the resistance. Similarly, as the Ohmic contact width is decreased from 'long' to 'point' to 'equal' there is an increase in resistance and a decrease in the capacitance. The change in resistance and capacitance with the geometry is consistent with the general formulae for capacitance (Eq. 2) and resistance (Eq. 3) shown below

$$R = \frac{\rho L}{A} \quad (2)$$

$$C = \frac{\varepsilon A}{d} \quad (3)$$

where $R$ is the resistance, $\rho$ is the resistivity of the material, $L$ is the length of the material, $A$ is the cross-sectional area of the material, $C$ is the capacitance and $d$ the distance between two parallel plates.

The frequency limit of the four sensors is measured using the setup in Fig 2 and plotted with respect to the geometric factor in Fig 5. As the geometric factor is increased, we see a drop in the frequency limit of the Hall effect sensor. The reduction can be attributed to an increase in the overall RC (resistance × capacitance) of the sensor, plotted as the theoretical frequency limit in Fig 5. Thus, the 'long' contact geometry has the highest frequency limit of 46.1 MHz, whereas the 'short' contact has the lowest frequency limit of 7.7 MHz. By extending the capacitance and resistance plots as a function of the geometric factor, we can extrapolate the frequency limit of the Hall effect sensor. As seen by the theoretical line for the frequency limit in Fig 5, there is a finite gain in the frequency limit when reducing the L/W$_{effective}$ geometry. By further reducing the L/W$_{effective}$, for example, by reducing the length of the active region (L), the capacitance of the Hall effect sensor increases exponentially while the resistance decreases linearly as seen in Fig 4. The increase in capacitance dominates the RC constant and thus caps the frequency limit of the device.

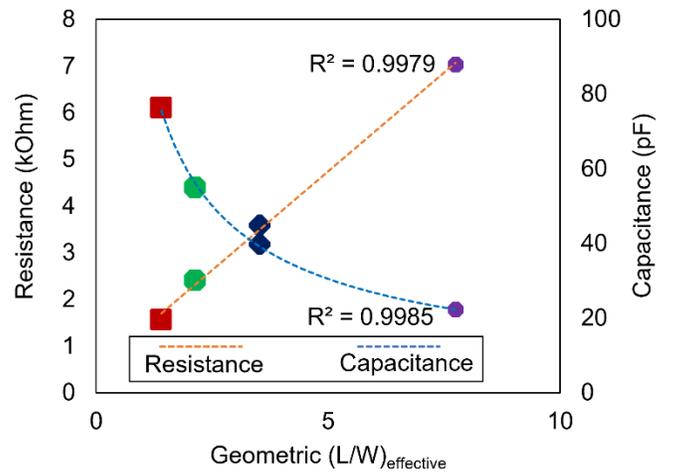

Fig. 4. Resistance and capacitance vs geometric factor for Hall effect sensor geometries. The resistance plot has a linear fit with an equation $y=0.85x+0.5$ and the capacitance plot has a power fit with the equation $y=96.5x^{-0.72}$.

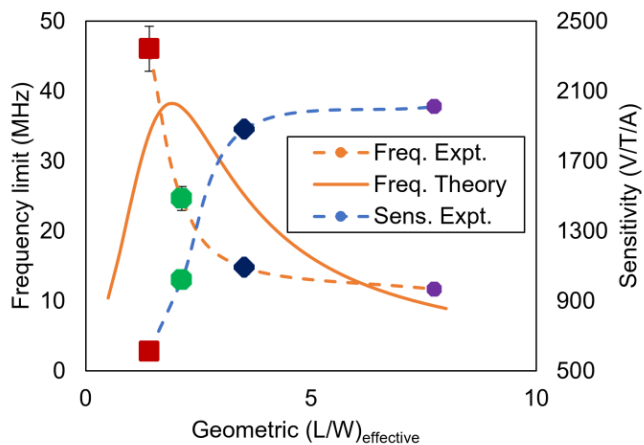

Fig. 5. Experimental results for the frequency limit, current scaled sensitivity, and calculated frequency limit from theory as a function of geometry.

In summary, there is a clear trade-off between current scaled sensitivity of the Hall effect sensor and the frequency limit based on the chosen geometry. To measure high frequency magnetic fields, such as in power electronics, there is a trade-off in sensitivity that must be made when choosing the geometry of the Hall effect sensor.

## IV. CONCLUSION

In this work, we establish the geometry dependent frequency limit for GaAs/AlGaAs based Hall effect sensors. Understanding the resistance and capacitance of the device gives insight into the practical limits for high frequency Hall effect sensors. The results are important for future Hall effect sensor designs where a trade-off between sensitivity and frequency needs to be made.

With the increased development of high frequency power electronics and their ability to go beyond the conventional Hall effect sensor frequency limit, capped by current spinning, it is exceedingly important to understand the frequency this class of sensors can achieve. High mobility devices such as 2DEG based GaAs/AlGaAs devices allow for higher frequency limits and further optimization of the geometry of the device can enable even higher frequency limits.

The temperature effect on the frequency limit for GaAs/AlGaAs and more temperature stable 2DEG-based systems will be presented in future work.

## V. ACKNOWLEDGEMENTS

The authors would like to thank Professor Gregory Salamo, Professor Alan Mantooth and Dr. Satish Shetty for their aid in discussions and overall planning of the manuscript, and Dr. Werner Schrenk for device processing discussions. This work was supported in part by the Stanford SystemX Alliance, and the National Science Foundation Engineering Research Center for Power Optimization of Electro-Thermal Systems (POETS) with cooperative agreements (NSF Award # 1449548), European Office of Aerospace Research and Development/Air Force Office of Scientific Research (EOARD/AFOSR FA8655-22-1-7170), Austrian Development Fund Green Sensing MIR (FFG 883941).

## References

[1] T. Chaouachi and L. Sbita, "The impact of Hall effect sensor filtered signals on the in-wheel motor tire of an electric vehicle," 2019 4th International Conference on Power Electronics and their Applications (ICPEA), Sep. 2019.

[2] M. Sen, I. Balabozov, I. Yatchev, and R. Ivanov, "Modelling of current sensor based on hall effect," 2017 15th International Conference on Electrical Machines, Drives and Power Systems (ELMA), Jun. 2017.

[3] A. Kumar and V. John, "Power electronic converter for characterization of Hall effect current sensors," 2014 IEEE International Conference on Power Electronics, Drives and Energy Systems (PEDES), Dec. 2014.

[4] A. Udo, "Limits of offset cancellation by the principle of spinning current Hall probe," SENSORS, 2004 IEEE, 2004, pp. 1117-1120 vol.3,.

[5] M. Crescentini, M. Biondi, R. Ramilli, P. A. Traverso, G. P. Gibiino, A. Romani, M. Tartagni, M. Marchesi, and R. Canegallo, "A broadband current sensor based on the X-hall architecture," 2019 26th IEEE International Conference on Electronics, Circuits and Systems (ICECS), Nov. 2019.

[6] A. Bilotti, G. Monreal, and R. Vig, "Monolithic magnetic hall sensor using dynamic quadrature offset cancellation," IEEE Journal of Solid-State Circuits, vol. 32, no. 6, pp. 829–836, 1997.

[7] A. Lalwani, A. S. Yalamarthy, D. Senesky, M. Holliday, and H. Alpert, "Hall effect sensor technique for no induced voltage in AC magnetic field measurements without current spinning," IEEE Sensors Journal, vol. 22, no. 2, 2022.

[8] Popović R. S., "4.6.1," in Hall effect devices, Bristol: Institute of Physics Pub., 2004, pp. 238–238.

[9] M. Crescentini, M. Marchesi, A. Romani, M. Tartagni, and P. A. Traverso, "Bandwidth limits in hall effect-based current sensors," ACTA IMEKO, vol. 6, no. 4, p. 17, 2017.

[10] H. S. Alpert, K. M. Dowling, C. A. Chapin, A. S. Yalamarthy, S. R. Benbrook, H. Kock, U. Ausserlechner, and D. G. Senesky, "Effect of geometry on sensitivity and offset of Algan/Gan and inaln/gan Hall effect sensors," IEEE Sensors Journal, vol. 19, no. 10, pp. 3640–3646, 2019.

[11] Y. Xu, A. Lalwani, K. Arora, Z. Yang, A. Renteria, D. G. Senesky, P. Wang, "Hall-Effect Sensor Design with Physics-informed Gaussian Process Modeling," in IEEE Sensors Journal, Oct. 2022

[12] M. L. M. da S. C. de. and I. Darwazeh, Introduction to linear circuit analysis and modelling from DC to RF. Amsterdam: Newnes, 2005.